\documentstyle{aipproc}

\input psfig

\def\xslide#1#2#3#4#5#6#7{\centerline{\psfig
{figure=#1,height=#2,bbllx=#3bp,bblly=#4bp,bburx=#5bp,bbury=#6bp,width=#7,clip=}}}

\newcommand{\be }{\begin{equation}}
\newcommand{\ee}{\end{equation}} 
\newcommand{\ba}{\begin{eqnarray}}
\newcommand{\ea}{\end{eqnarray}} 
\newcommand{\pp}{$\pi\pi$ }
\newcommand{\pipi}{$\pi^+\pi^-$ }
\newcommand{\KK}{$K\overline{K}$ } 
\newcommand{\fo}{$f_0(980)$ }
\newcommand{\ro}{$\rho(770)$ } 
\newcommand{\ftens}{$f_2(1270)$ }
\newcommand{\rf}{$\rho_3(1690)$ } 

\newcommand{\aone}{$a_1$ }

\newcommand{\mpp}{$m_{\pi\pi}$ }

\newcommand{\del}{$\delta$ }
\newcommand{\et}{$\eta$ } 

\newcommand{\downp}{"down--flat" } 
\newcommand{\downs}{"down--steep" }
\newcommand{\upp}{"up--flat" } 
\newcommand{\ups}{"up--steep" }

\newcommand{\reactpol}{\mbox{$\pi^- p_{\uparrow} \rightarrow \pi^+ \pi^- n$} }
\newcommand{\reactnonpol}{\mbox{$\pi^- p \rightarrow \pi^+ \pi^- n$} }

\begin{document}

\title{New Solutions for Scalar--Isoscalar \pp Phase Shifts}

\vspace{-0.3cm}

\author{R.  Kami\'nski, L.  Le\'sniak and
K. Rybicki}

\vspace{-0.5cm}

\address{Henryk Niewodnicza\'nski Institute of Nuclear Physics, \\
PL 31-342 Krak\'ow, Poland}

\maketitle

\vspace{-1cm}
 
\begin{abstract}
The scalar--isoscalar $\pi\pi$ phase shifts are calculated 
in the \pp energy range from 600 MeV to 1600 MeV.
We use results of the CERN--Cracow--Munich collaboration for the reaction 
\reactpol
on a transversely polarized target at 17.2 GeV/c $\pi^-$ momentum. 
Energy--independent separation of the $S$--wave pseudoscalar amplitude 
($\pi$ exchange) from the pseudovector amplitude (\aone exchange) is
carried out. 
Below the $K\overline{K}$ threshold we find two solutions for the 
\pp phase shifts, for which the phases increase slower with the
effective \pp mass than the P-wave phases ("flat" solutions)
and two solutions for which the phases increase faster than the P-wave phases
("steep" solutions).
Above 1420 MeV both sets of phase shifts increase with energy
faster than in the experiment on an unpolarized target.  This fact can
be related to a presence of the scalar resonance $f_0(1500)$.

\end{abstract}


\vspace{-0.5cm}

One of the main sources of information on the scalar-isoscalar 
\pp interactions  is the \pipi partial wave analysis (PWA) yielding the 
$S$--wave.
Virtually all PWA's were based on the old CERN--Munich experiment 
[1] 
for the reaction \reactnonpol on nonpolarized target at 17.2 GeV/c.
The number of observables provided by such experiment is much smaller
than the number of real parameters needed to describe the partial waves.
Consequently, the dominance of pseudoscalar exchange, equivalent to the
absence of pseudovector exchange and several other physical assumptions
have been made in previous studies [1--6].

In this analysis we use results of PWA performed by
the CERN--Cracow--Munich collaboration for reaction
$\pi^-p_{\uparrow}\rightarrow \pi^+\pi^- n$ on polarized target 
at 17.2 GeV/c
[7].
The data cover the \pp effective mass range from 
600 MeV to 1600 MeV and the $t$--momentum transfer range from
$-0.2$ (GeV/c)$^2$ to $-0.005$ (GeV/c)$^2$.

Combination of results of experiments on polarized and nonpolarized target 
yields a number of observables sufficient for performing a quasi--complete 
and energy independent PWA without any model assumptions.  
This analysis is only quasi--complete because of an unknown phase between 
two sets of transversity amplitudes.  
Nevertheless, intensities of partial waves could be determined in a completely 
model--independent way.  
This removed ambiguities appearing in earlier studies, except for the old
"up--down" ambiguity 
[2]. 
General belief (see e.g.  
[8,9]) 
was that the "up--down" ambiguity had been resolved definitely in favour 
of the "down" solution.  
We disagree with this belief since all the studies of the
{\it full} (i.e. including polarized-target) data are consistent with both the
``up'' and ``down'' solutions.


Let us denote by $f_0$ a system of two pions in a relative
$S$--wave isospin 0 state.  Transition amplitude for the $f_0$
production process $\pi^{-}p\rightarrow f_0~n$ can be written as the
following matrix element 

\vspace{-0.3cm}

\be  T_{s_p s_n}=<u^{s_n}_{p_2}|A\gamma_{5}+
\frac{1}{2}B\gamma_{5} \gamma_{\mu}(p_{\pi}+p_f)^{\mu}|u^{s_p}_{p_1}>,
\label{1} \ee where $p_1,p_2,p_{\pi}$ and $p_f$ are proton, neutron,
incoming pion, and final $f_0$ four-momenta, $s_p$ and $s_n$ are proton
and neutron spin projections, $u^{s_p}_{p_1}$ and $u^{s_n}_{p_2}$ are
the corresponding four-spinors, A and B are functions of the Mandelstam
variables $s=(p_1+p_{\pi})^2$ and $t=(p_1-p_2)^2$ at fixed $f_0$ mass
$m_{\pi\pi}$.  Part A of the amplitude corresponds to the pseudoscalar
(or one pion) exchange while part B describes the pseudovector exchange
which we shall briefly call $a_1$ exchange.
Functions A and B have to be determined from experiment.  

In this paper we use two transversity amplitudes g and h, adequate for 
describing $f_0$ production on a transversely polarized target:
\mbox{$g\equiv<n\downarrow|T|p\uparrow>$} and
\mbox{$h\equiv<n\uparrow|T|p\downarrow>$}.
Separation of the amplitudes $g$ and $h$ into two components
($g=g_A+g_B$ and $h = h_A+h_B$) proportional to the invariant amplitudes
$A$ and $B$ has been described in
[10].
The amplitudes $g_i$ and $h_i$ ($i=A,B$) have been 
averaged over $t$.

Evaluation of the amplitude $A$ allows us to describe 
the $S$--wave $\pi^+\pi^-\rightarrow \pi^+\pi^-$ elastic scattering 
amplitude $a_S$ in the following way:  


\be  a_S=-
\frac{p_{\pi}\sqrt{sq_{\pi}}\,f}{m_{\pi\pi}\sqrt{2\cdot\:
\frac{g^2}{4\pi}}}
\frac{A(m_{\pi\pi})}{2M}, \label{38} \ee where $p_{\pi}$ is the incoming $\pi^-$
momentum in the $\pi^-p$ c.m.  frame, $q_{\pi}$ is the final pion
momentum in the $f_0$ decay frame, $g^2/4\pi=14.6$ is the pion-nucleon
coupling constant, and $f$ is the correction factor (see [10]).  


Scalar-isoscalar pion-pion amplitude $a_0$ can be expressed as a function
of $a_S$ and the $I$=2 $S$--wave amplitude $a_2$ and normalized to 
Argand's form: 
\be  
a_0=3~a_S-\frac{1}{2}a_2 = \frac{\eta e^{2i\delta}-1}{2~i},
\label{42} 
\ee 
where $\delta$ is the $I$=0, $S$=0 \pp phase shift and $\eta$
is the inelasticity coefficient.

In our analysis we have assumed that phases of the $P$, $D$
and $F$--waves follow phases of \ro,  \ftens and \rf decay amplitudes into
\pipi.
In the region 600 MeV -- 980 MeV we assume that only the $S$ and $P$--waves
contribute. It is only in this region that fully analytical solutions of
the PWA equations are possible [11].  The PWA analysis however,
yields two solutions ("up" and "down") which are distinctly different in
the \mpp region from 800 MeV to 980 MeV.
In addition to the
"up--down" ambiguity in the moduli of the $g$ and $h$ transversity
amplitudes, there is also a phase ambiguity in each \mpp bin.  This
ambiguity comes from the mathematical structure of the PWA equations
from which only cosines of the relative phases of the partial waves 
can be obtained.  
In our analysis we present two arbitrary choices of the $S$--wave phases.  
In the first set, called "steep", $S$--wave
phases grow faster than $P$--wave phases. 
In the other set, called "flat", increase in $S$--wave phases is slower 
than that for $P$--waves.  
Thus two sets of possible phases (``flat'' or ``steep'') combined
with two branches of moduli (``up'' and ``down'') lead us to four solutions.
Let us remark that in [2] the words "up" and "down" serve to
distinguish the sets of phases while in our case they serve to distinguish the
sets of moduli.  

\begin{figure}[h!] 
\vspace{-0.3cm}

\xslide{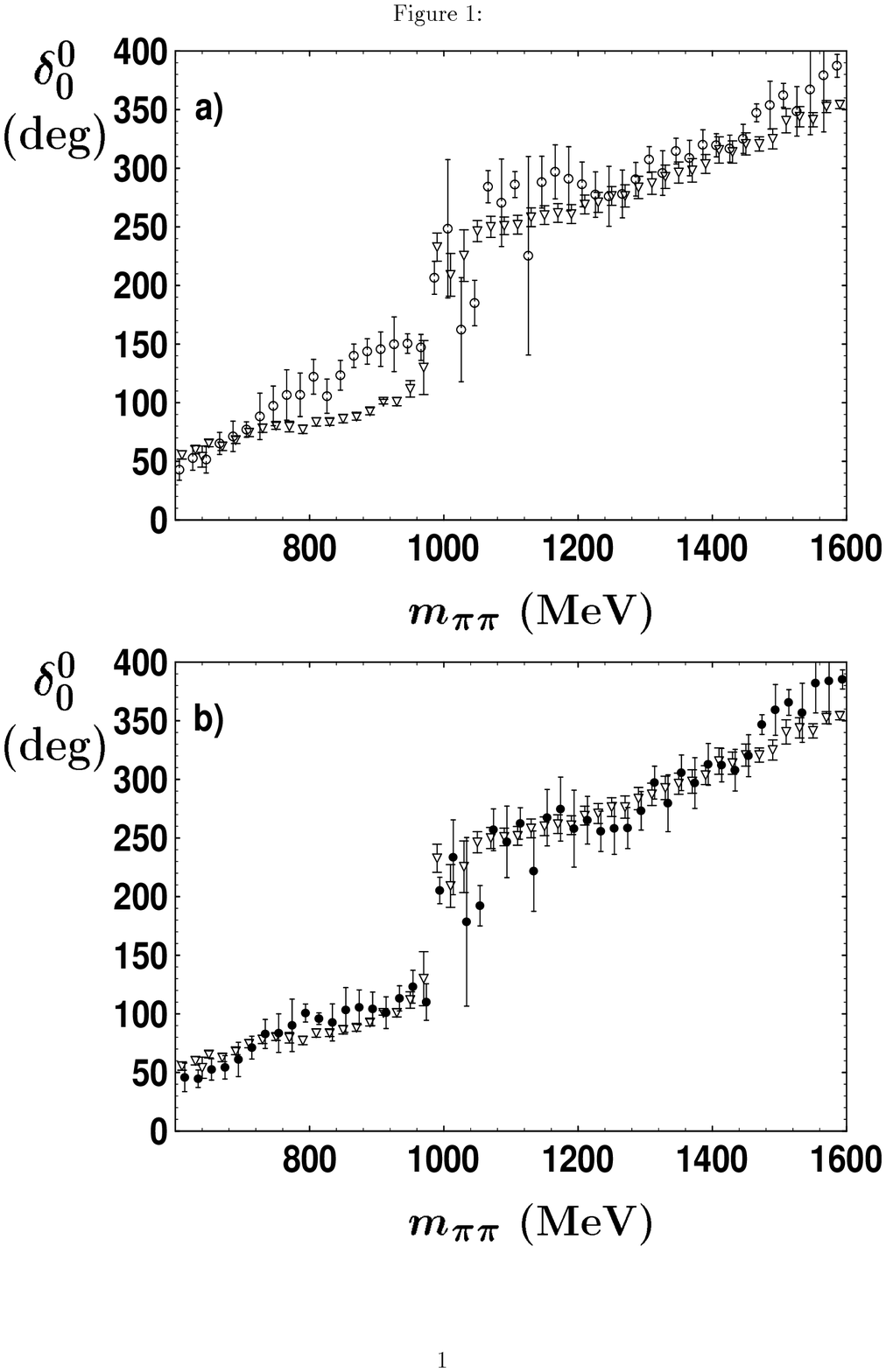}{7cm}{-370}{95}{525}{410}{14.2cm}
\vspace{-7cm}
\hspace{-3.6cm}\xslide{fig1.ps}{7cm}{69}{410}{525}{720}{7.2cm}
\vspace{0.5cm}
\xslide{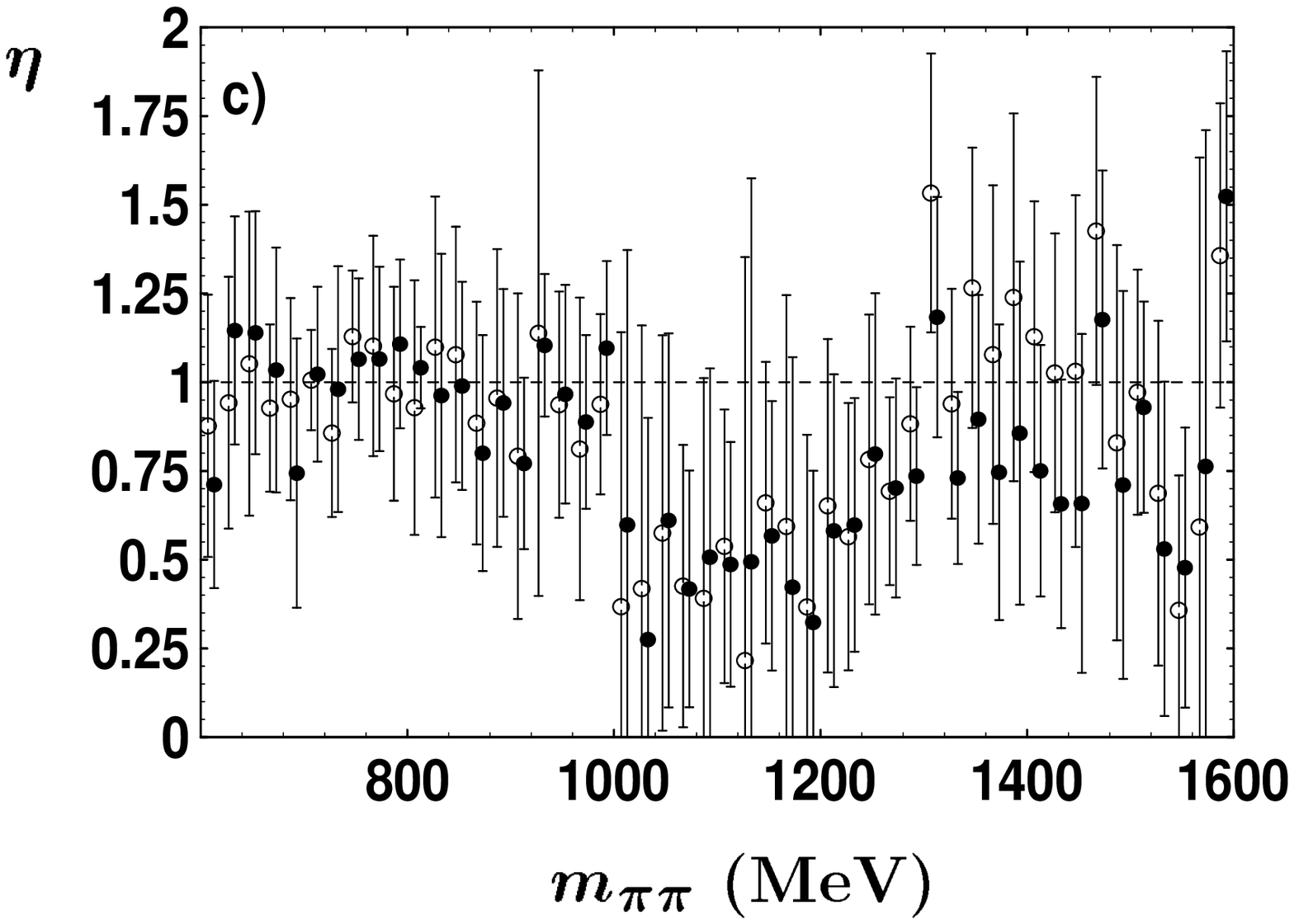}{7cm}{45}{145}{535}{495}{8cm} 
\vspace{-0.5cm}
\caption{\small {\bf a)} Scalar--isoscalar \pipi
phase shifts $\delta_0^0$ as a function of the effective \pipi mass for
the \upp solution (open circles) and for data [1] (triangles).  {\bf
b)} Same as in a) for the \downp solution (full circles).  {\bf c)}
Scalar--isoscalar \pipi inelasticity coefficient \et versus the
effective \pipi mass for the \upp (open circles) and \downp (full
circles) solutions.}
\label{fig_down}
\end{figure}

Phase shifts \del shown in Fig. \ref{fig_down} a,b have been compared 
with those obtained from the analysis of the \reactnonpol
reaction on the  unpolarized target 
[1] (solution B),
where separation of the $\pi$ and \aone exchanges was impossible.  
In Fig. \ref{fig_down}a in the mass region from 600 MeV to about 1400 MeV 
we do not see any systematic differences between phase shifts corresponding 
to the \downp solution and the results of [1].  
In the \upp solution around 900 MeV, however, the values of \del are higher 
than those measured in [1] by several tens of degrees (Fig. 1b).
As it can be seen in Fig.
\ref{fig_down}c, inelasticity is close to 1 up to about 1000 MeV.  
A sudden drop in \et for the effective mass near 1000 MeV along with a 
characteristic jump of phase shift \del is caused by 
opening of a new \KK channel and presence of the narrow \fo resonance 
[12,13].  
Another decrease of \et near 1500 MeV -- 1600 MeV can be related to the
opening of further channels like $4\pi$ ( $\sigma\sigma$ or $\rho\rho$), 
$\eta\eta'$ or $\omega\omega$.  

For two solutions \ups and \downs (not shown in Fig. 1c) 
we observe a characteristic fast increase of phase shifts near 780 MeV
(see [10]).
We see also rapid changes of \et 
which exceeds substantially 1 for two regions
around 600 MeV -- 700 MeV and 830 MeV -- 980 MeV (especially in the
``down-steep'' solution). 
We conclude that this solution is unphysical and 
can be ignored, however the ``up-steep'' solution, although a bit peculiar, 
cannot be completely excluded.

We would like to point out that for 
\mpp larger than 1470 MeV, in all our solutions (``flat'' and ``steep'') 
the phase shifts 
show a systematically
steeper increase than the phase shifts corresponding to data obtained on the
unpolarized target. 
This increase is related to a presence of the relatively narrow 
(90 MeV - 170 MeV)
resonance with mass 1400 MeV - 1460 MeV (see refs. 
[14,15]).

\vspace{0.1cm}

This work has been partially supported by the Polish State Committee
for Scientific Research (grants No 2P03B 231 08 and 2 P03B 043 09).


\end{document}